\documentclass{appolb}
\usepackage{epsfig}
\usepackage{here}
\begin{document}
\title{STUDY OF THE NEAR THRESHOLD TOTAL CROSS SECTIONS 
       FOR THE QUASI-FREE $pn \to pn \eta^{\prime}$ REACTION 
       WITH THE COSY-11 DETECTOR%
\thanks{Presented at the Symposium on Meson Physics, Cracow, 01-04 October 2008.}%
}
\author{Joanna~Klaja$^{\star}$, Pawe{\l}~Moskal$^{\star,\$}$, Jaros{\l}aw~Zdebik$^{\star}$\\
on behalf of the COSY--11 collaboration
\address{
$^{\star}$Institute of Physics, Jagellonian University, Cracow, Poland\\
$^{\$}$Institute for Nuclear Physics and J{\"u}lich Center for Hadron Physics, \\
Research Center J{\"u}lich, Germany}
}
\maketitle
\begin{abstract}
The measurement of the quasi-free production of the
$\eta^{\prime}$ meson has been carried out at the COSY--11 detection
setup using a proton beam and a deuteron cluster target.
The energy dependence of the cross section is extracted using a fixed
proton beam momentum of
$p_{beam}=3.35$~GeV/c
and exploiting the Fermi
momenta of nucleons inside deuterons. The data cover a range of centre-of-mass
excess energies from 0 to 24~MeV. Due to the low statistics 
at the present stage only upper limits for the cross sections could be extracted.
In this article we focus 
on the functioning and efficiency determination of the neutron detector. 

\end{abstract}
  
\section{Introduction}

In August 2004 --for the first time-- using the COSY--11~\cite{cosy11} facility
we have conducted a measurement of the $\eta'$ meson production
in the proton-neutron collision~\cite{moskal01}.
The aim of the experiment is the determination of the total cross section of the
$pn\to pn \eta^{\prime}$ reaction near the kinematical threshold.
We expect that the comparison of the
$pp \to pp\eta'$ ~\cite{ppetap} and $pn \to pn\eta'$ total cross sections
will help to understand  the production of the $\eta'$ meson in
different isospin channels and to investigate aspects of the
gluonium component of this meson.
In the framework of the quark model the $\eta^{\prime}$ meson is predominantly
a flavour-singlet combination of quark-antiquark pairs, and it is expected
to mix with purely gluonic states.
Therefore, additionally to the production mechanisms associated with meson
exchange~\cite{nakayama01,kampfer01}
it is also possible that the $\eta^{\prime}$ meson is produced from excited glue
in the interaction region of the colliding nucleons, which couple to the $\eta^{\prime}$
meson directly via its gluonic component or through its SU(3)-flavour-singlet
admixture~\cite{bass02,bass03}. 

\section{Experimental method}

In order to investigate the isospin dependence of the meson production
in hadronic interactions~\cite{moskal-prog-part}, the COSY--11
facility has been extended by a neutral--particle--detector.
The measurement of the $\eta^{\prime}$ meson production in the proton-neutron 
collision~\cite{moskal01} has been conducted using a proton beam of the Cooler Synchrotron
COSY~\cite{cosy} and a cluster jet deuteron target~\cite{dombrowski}.
More details concerning functioning of the particular detectors and 
method of the measurement of the quasi-free reaction
can be found in references~\cite{cosy11,moskal01,przerwa01,jklaja01}.
In this article  we will concentrate on the neutral particle 
detector~\cite{czyzykpraca,rozek-raport,przerwa} 
which has been designed to deliver the time at which the registered
neutron or gamma quantum induced a hadronic or electromagnetic reaction,
respectively. This information combined with the time of the reaction
at the target place --- deduced using other detectors --- enables to calculate
the time-of-flight between the target and the neutron detector
and to determine the absolute momentum of registered paricles,
provided that they could have been identified.\\
\begin{figure}[H]
{\centerline{\includegraphics[height=.26\textheight]{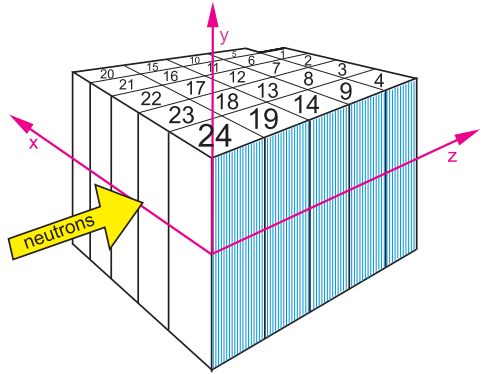}}}
\caption{Configuration of the detection units}
\label{neut_3d}
\end{figure}
The neutron detector consists of 24 modules.
Each module is built out of eleven plates of
scintillator material with dimensions of 240~mm $\times$ 90~mm $\times$ 4~mm interlaced with
eleven plates of lead with the same dimensions. The scintillators are read out
at both edges of the module via light guides --- made of plexiglass --- whose
shape changes from rectangular to cylindrical, in order to accumulate the
produced light on the circular photocathode of a photomultiplier.
The neutron detector is positioned at a distance
of 7.36~m
from the target with the configuration of modules schematically depicted
in figure~\ref{neut_3d}. The detector covers the neutron laboratory
angular range of $\pm 1.84^o$ in x and $\pm 1.1^o$ in y direction.

\begin{figure}[H]
\includegraphics[height=.29\textheight]{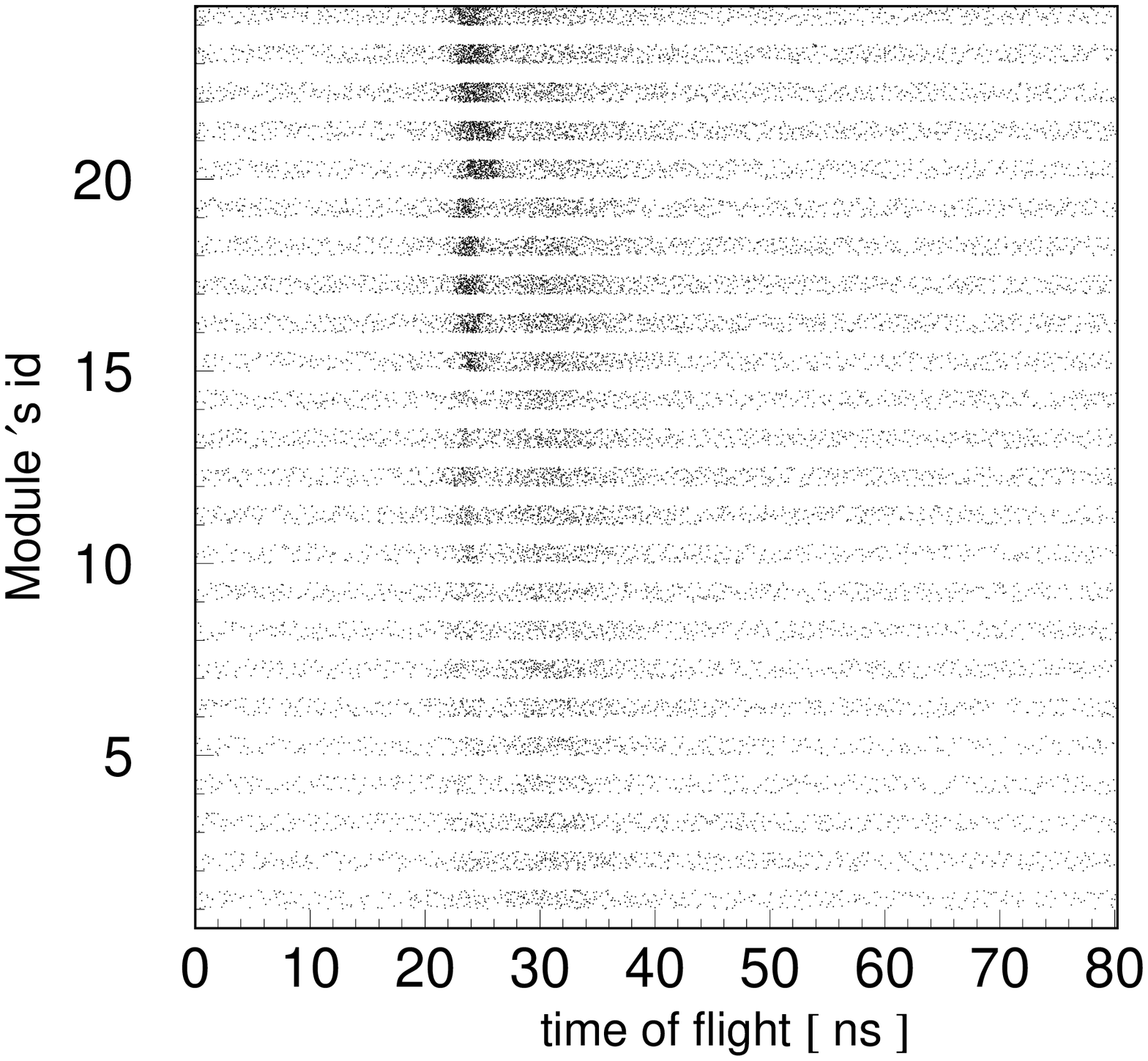}
\includegraphics[height=.29\textheight]{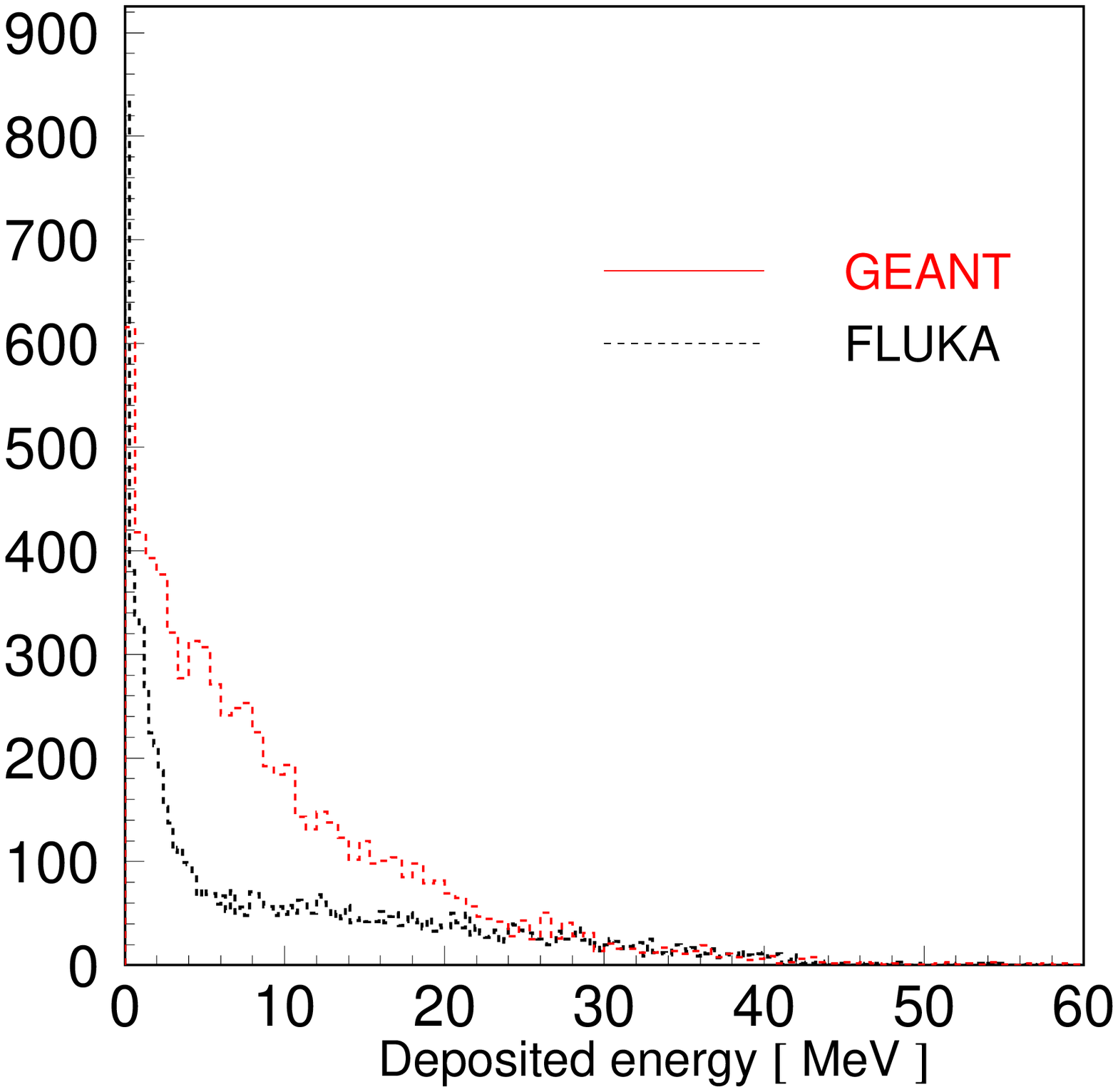}
\caption{ {\bf Left}: Time--of--flight determined between the target and
                      the neutral particle detector  for all detector modules.
         {\bf Right}: Total energy deposited in scintillator layers of the neutral particle detector
                      for neutrons impinging on the detector with kinetic energy 
                      equal to 300~MeV from simulations using FLUKA-2008 (dashed line) 
                      and GEANT-3 (dotted line) packages.}
\label{neut}
\end{figure}

Figure~\ref{neut} (left) shows the distribution of the time--of--flight
for all detection units. One can
see a clear enhancement of events originating from gamma quanta.
As expected from the known absorption coefficients~\cite{Hagiwara},
gamma quanta are predominantly registered in the first row of the
detector whereas interactions
points of neutrons are distributed more homogeneously.\\
%--------------------------------------------
%   Czesc dotyczaca efficiency              |
%--------------------------------------------
The efficiency of the COSY--11 neutral particle detector --
which is an important factor for determining the absolute 
values of cross sections -- was determined using two independent
simulation programs. One of these programs is based on the
GEANT-3 ({\bf GE}ometry {\bf AN}d {\bf T}racking) code~\cite{geant} 
used for simulation of the hadronic cascades induced in matter by neutrons.
In the other program the FLUKA\footnote{The simulations were
performed with the new 2008 version.}
({\bf FLU}ktuierende {\bf KA}skade) simulation package is used~\cite{fluka,zdebik01}.

\begin{figure}[H]
\includegraphics[height=.29\textheight]{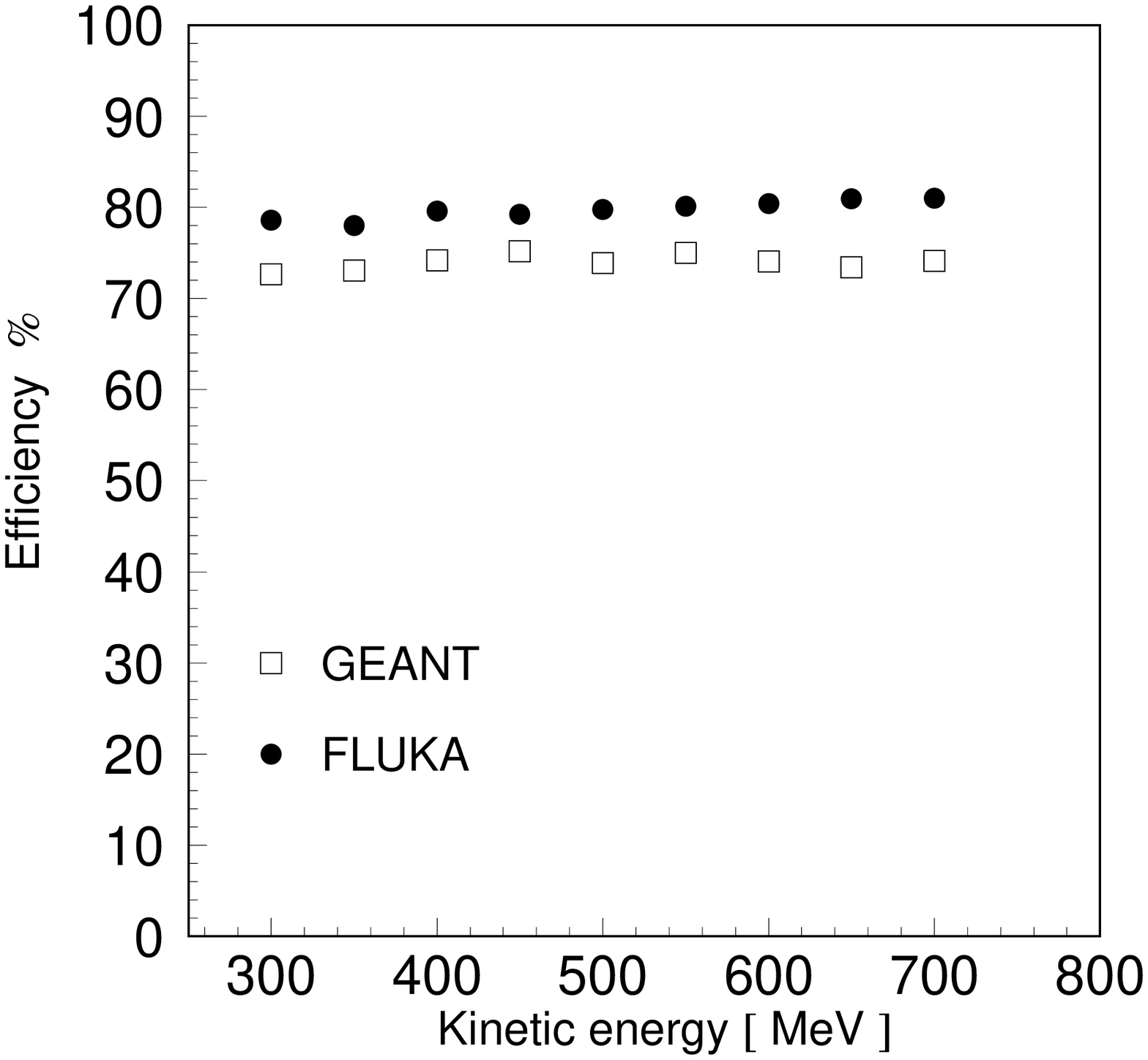}
\includegraphics[height=.29\textheight]{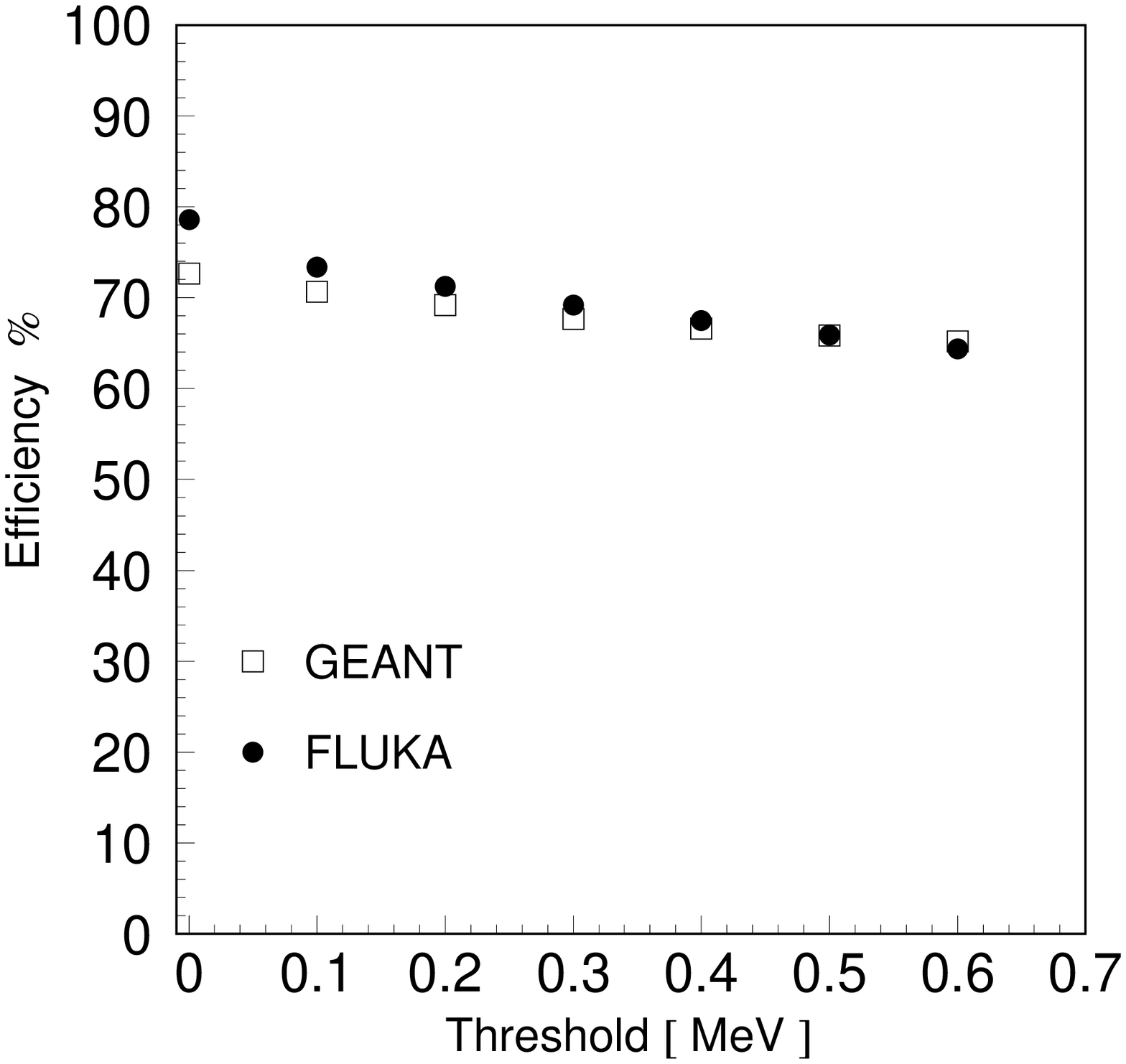}
\caption{ {\bf Left}: The efficiency distribution as a function of the kinetic energy
                      of neutrons determined for a  threshold of 0 MeV.
         {\bf Right}: Relation between the threshold value and the efficiency for 
                      neutrons with energy of 300 MeV.}
\label{fluka_geant}
\end{figure}

The comparison of the simulated distributions of the total deposited energy
in the neutron detector using FLUKA-2008 and the GEANT-3 packages, is presented
in fig.~\ref{neut}.
The simulation was performed for neutrons
with kinetic energy equal to 300 MeV. As can be seen, the range of
deposited energy is the same for both cases, however, GEANT simulations
yield on the average higher energy response of the neutron detector.
The efficiency of the neutron detector is given by the ratio of the number of generated neutrons
to the number of events, for which the  energy deposited in the scintillator material was
larger than the threshold value in at least one of 24 detection units.
The calculated efficiency as a function of the kinetic energy of neutrons
is shown in fig.~\ref{fluka_geant} (left). In this figure open squares denote results obtained
using the GEANT-3 package, and  the outcome of simulations using the FLUKA-2008 is presented as  black
circles.

The kinetic energy of neutrons from the $pn \to pn \eta^{\prime}$ 
reaction varies from 300~MeV up to 700~MeV for the 3.35~GeV/c beam 
momentum, and as can be inspected from fig.~\ref{fluka_geant} (left) the 
efficiency is fairly constant in this range. 
It is worth to stress that two independent
simulation tools leads to fairly good ($\pm3\%$ ) agreement
for the values of the efficiency in the energy range  relevant for 
the studies of the $pn \to pn \eta^{\prime}$ reaction. \\
We have also conducted studies of the efficiency  dependency 
on the threshold. In the experiment the 
threshold was set to about 0.1 MeV and therefore
we scanned the values from 0 up to 0.6~MeV.
The result is presented in fig.~\ref{fluka_geant} (right).
For both the GEANT and FLUKA-2008 simulations 
the values of the efficiency change 
by about 10\% over the scanned threshold range of 0.6~MeV.

%--------------------------------------
%   Koniec dzialu efficiency.         |
%--------------------------------------
\subsection{Estimation of the total cross section}

Having all outgoing nucleons from $pn \to pn\eta^{\prime}$ reaction registered,
we apply the missing mass technique in order to identify the $\eta^{\prime}$
meson. Due to the smaller efficiency and lower resolution for the registration
of the quasi-free $pn \to pn~meson$
reaction in comparison to the measurements of the proton-proton reactions,
the elaboration of the data encounters problems of low statistics.
Therefore, the excess energy range for $Q \ge 0$ has been divided
only into four intervals of 8~MeV width. For each interval
we have calculated the missing mass. Next, from events with negative Q value
the  corresponding background missing mass
spectrum was constructed, shifted to the kinematical limit 
and normalized to the experimental distribution
at low mass values where no events of the $\eta^{\prime}$ meson production are expected.
A detailed description of the method used for the background subtraction
can be found in  a dedicated article~\cite{moskal02}.
After subtracting missing mass distributions for the negative values of Q 
from spectra for Q values larger than 0 --
due to the very low signal--to--background ratio --
at the present stage of the data analysis,
the signal from the $\eta^{\prime}$ meson was found to be statistically insignificant.
\begin{figure}[H]
{\centerline{\includegraphics[height=.29\textheight]{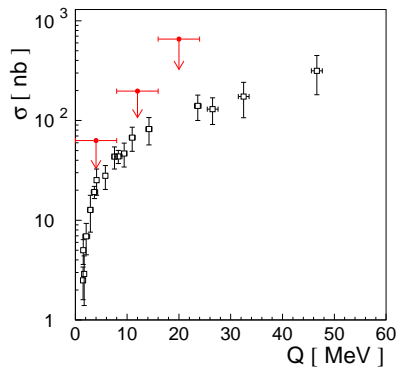}}}
\caption{Total cross sections for the $pp \to pp\eta^{\prime}$
         reaction as a function of the excess energy (open symbols).
         Upper limit for the total cross section for the $pn \to pn\eta^{\prime}$
         reaction as a function of the excess energy (closed symbols). }
\label{result}
\end{figure}

Nevertheless, having the luminosity  --~established from
the number of the quasi-free
proton-proton elastic scattering events~\cite{aiplumi}~--
and the detection efficiency of the COSY-11 system we have estimated
upper limits of the total cross section for the quasi-free
$pn \to pn \eta^{\prime}$ reaction. The preliminary result is shown
in Fig.~\ref{result}.

\section{Perspectives}
As already mentioned based on the $pp\to pp\eta^{\prime}$ reaction channel only  
it is not possible to determine the reaction machanism responsible for the $\eta^{\prime}$ production.
To put more constrains on the theoretical models~\cite{nakayama01,kampfer01} and to learn more 
about the structure of the $\eta^{\prime}$ meson~\cite{bass02,bass03} we began investigations
aiming at the determination of the dependence of the $\eta^{\prime}$ meson production
on the isospin of the colliding nucleons.
After the measurement of the $pn\to pn\eta^{\prime}$ reaction, we have extended our experimental
studies to a pure isospin zero state of the interacting nucleons by the measurement
of the quasi-free $pn\to d\eta^{\prime}$ reaction. This experiment was conducted using
a proton beam with a momentum of 3.365~GeV/c and a deuteron target.
Assuming that the ratio of the cross sections for the $pn\to d\eta^{\prime}$
and $pp\to pp\eta^{\prime}$ reactions will be at the same order as the ratio already 
established~\cite{calen02} for the $pn\to d\eta$ and $pp\to pp\eta$ reactions we expect
to identify about 1000 $pn\to d\eta^{\prime}$ events in the available data sample~\cite{basia}.\\
An additional outcome of this measurement will be an increase of the statistics for the
$pn\to pn\eta^{\prime}$ channel and a consistency check of the obtained results.\\
Together with the previous results on $pp\to pp\eta^{\prime}$ and $pn\to pn\eta^{\prime}$
we would then complete the study of the
$\eta^{\prime}$ meson production cross section in nucleon-nucleon collisions.
The results obtained in different isospin channels
can then be compared with theoretical models
for the production of mesons. This will
reduce significantly the ambiguities of such models and will lead to a better understanding
of the production mechanism of the $\eta^{\prime}$ meson in nucleon-nucleon collisions.
The result will be also of importance in rate estimates 
for studying the $\eta^{\prime}$ meson decays
with the WASA-at-COSY facility~\cite{hhadam01,mzielinski01}.\\

\section{Acknowledgments}
We acknowledge the support by the
European Community
under the FP6 programme (Hadron Physics,
RII3-CT-2004-506078), by
the German Research Foundation (DFG) and by 
the Polish Ministry of Science and Higher Education under grants
No. 3240/H03/2006/31, 1202/DFG/2007/03, 0082/B/H03/2008/34.

\end{document}